\documentstyle[12pt,epsfig,amsmath,amssymb,pstricks]{article}
\begin{document}

\title{\bf Autonomy all the way down: \\Systems and dynamics in \\quantum Bayesianism}

\author{{Chris Fields}\\ \\
{\it 21 Rue les Lavandi\`eres}\\
{\it Caunes Minervois 11160 France}\\ \\
{fieldsres@gmail.com}}
\maketitle

\begin{abstract}
Quantum Bayesianism (``QBism'') has been put forward as an approach to quantum theory that avoids foundational problems by altogether disavowing the objective existence of quantum states.  It is shown that QBism suffers its own versions of the familiar foundational problems, and that these QBist versions are illuminating not just for QBism, but for more traditional foundational approaches as well.
\end{abstract}


\textbf{PACSS:} 03.65.Ca; 03.65.Ta; 03.65.Yz
%

\section{Introduction}

It is rare for a program overview in theoretical physics to be deeply philosophical, to offer a broadly-based, sustained meditation about what the formal results that have been and are hoped to be obtained might mean.  Christopher Fuchs has given us such a paper: ``QBism: The perimeter of quantum Bayesianism'' (\cite{fuchs:10}; unless otherwise indicated, all quotations are from this paper).  Fuchs defines QBism as the view that ``quantum theory is not something outside probability theory ... but rather it is an \textit{addition} to probability theory'' (p. 9; emphasis in original).  Specifically, quantum theory is proposed to be ``an addition to personal, Bayesian, normative probability theory'' (p. 19), the particular variant of probability theory in which probabilities are \textit{individual} ``degrees of belief'' and are not in any sense objective facts about the world outside of some individual reasoning agent's head.  For the QBist, quantum states are not objective states of the world, and quantum information is not objective information about the world: quantum states are (just) quantum information, and quantum information is information for a specific, individual observer about ``the consequences (for \textit{me}) of \textit{my} actions upon the physical system'' (p. 7; emphasis in original).  While Fuchs derides ``stale, lifeless philosophy'' (p. 10) when advertising the theorem-proving prowess of the QBist movement, ``QBism: The perimeter of quantum Bayesianism'' is straightforwardly philosophical, embracing such ``false-started philosophies of 100 years ago'' (p. 1, Abstract) as meliorism and pragmatism, quoting William James almost as extensively as Einstein, and concluding that QBism ``hints of a world, a pluriverse, that consists of an all-pervasive `pure experience' '' (p. 27).  

While the scope of ``QBism: The perimeter of quantum Bayesianism'' is quite broad, two principal themes stand out.  The first is the role of measurement interactions, implemented by positive operator-valued measures (POVMs; for definitions, see \cite{fuchs:02}, Sect. 4 or \cite{nielsen-chaung:00}, Ch. 2), in updating observers' beliefs about the world.  The second is an elaboration of Fuchs' proposal, first made in \cite{fuchs:02}, that the \textit{physical}, as opposed to purely mathematical, content of quantum theory concerns the influence of the ``dimension'' $d$, ``formerly Hilbert-space dimension'' (p. 1, Abstract), of a physical system on its observable behavior.  These themes provide a realist counterpoint to the instrumentalism about quantum states that is the most well-known and widely-discussed feature of QBism \cite{caves:02a, caves:02b, caves:07, palge:08, timpson:08, friederich:11, fuchs:11a}.  Fuchs discusses the metaphysical and ontological questions raised by these realist themes ``with special emphasis on the view's outer edges'' (p. 1, Abstract), and in terms that are considerably more philosophically pugnacious than those employed even in his own earlier writings.  He provides his own versions of a number of arguments also made by others, defending QBism from charges of solipsism, resolving the question of how apparatus are to be defined in QBism by treating them as prosthetic extensions of observers, and outlining a decidedly non-instrumentalist QBist account of scientific explanation.  Perhaps in response to the proposal of Timpson \cite{timpson:08} that QBism requires an ontology along the anti-determinist lines of Cartwright's \cite{cartwright:99}, Fuchs emphasizes the objective indeterminism of the QBist universe, and employs the terms ``autonomous'' and ``creation'' to characterize quantum systems and processes, respectively.  Perhaps in response to Timpson's argument that QBist anti-determinism nonetheless does not entail a Jamesian pragmatism, Fuchs explicitly endorses James' position, an endorsement foreshadowed by his earlier remark that ``I should work a little harder to make myself look Jamesian'' (\cite{fuchs:02}, Acknowledgments).  To assure that no one doubts the seriousness of his most significant ontological claim, he puts it in capitals: ``QUANTUM STATES DO NOT EXIST'' (p. 2; emphasis in original).  It is surely unfair to attribute to all instrumentalists about quantum states or even to all quantum Bayesians the particular philosophical positions advanced in ``QBism: The perimeter of quantum Bayesianism.''  Timpson, for example, advances a reconstruction of quantum Bayesianism that is ``far more conservative than the sort of position I take it he (Fuchs) would most prefer'' (\cite{timpson:08}, p. 595).  Bub proposes that ``a quantum theory is fundamentally a theory about the possibilities and impossibilities of information transfer in our world ... not a theory about the mechanics of nonclassical waves or particles'' (\cite{bub:04}; p. 20), but counsels agnosticism in place of metaphysical or ontological speculation as a result.  However, Fuchs' arguments for QBism are compelling and the position sketched is nothing if not radical; it is therefore worth examining \textit{as philosophy}, and searching out the physical, metaphysical and ontological assumptions underlying it.

Fuchs begins ``QBism: The perimeter of quantum Bayesianism'' with a metaphor: like an annual bout of the flu, the foundational problems of quantum mechanics are a ``common annoyance'' that many physicists fear ``may one day turn out to be a symptom of something fatal'' (p. 1; see also \cite{fuchs:02}).  Fortunately, QBism provides a vaccine: ``To take this substance into one's mindset is all the vaccination one needs against the threat that quantum theory contains something viral for theoretical physics as a whole'' (p. 9).  The present paper closely examines the ingredients of Fuchs' vaccine, and shows that it in fact functions only as an analgesic, one with strange but revealing side effects.  QBism suffers its own versions of the standard foundational difficulties, versions from which new insights for quantum theory as a whole may be obtained.  As will be shown, these QBist versions of the standard foundational difficulties enter the theory through its two realist assumptions: the assumption that interactions with the world cause experiences and hence the revision of beliefs, and the assumption that physical systems have objective ``dimensions.''  A close comparison with quantum Darwinism and the ``environment as witness'' formulation of decoherence theory on which it rests \cite{zurek:98rough, zurek:03rev, zurek:04, zurek:05, zurek:06, zurek:07grand, zurek:09rev} is used throughout to illustrate both the novelty and the familiarity of QBism's underlying assumptions, and to show by example how the more problematic consequences of these assumptions can arise within interpretative approaches very different from QBism.

The ingredients of the QBist vaccine are assessed in three stages.  The first examines the metaphysics implicit in Fuchs' characterization of measurement interactions between an observer and an external quantum system.  Two consequences of this metaphysics that Fuchs closely approaches but leaves unstated are made explicit: that the QBist world has no objective dynamics, and that all physical systems are, for Fuchs, autonomous agents.   Hence the QBist view is ``autonomy all the way down,'' a metaphysical stance that makes the ``horror of non-locality'' perhaps a bit less horrible by comparison.  The second stage critically examines the ontology that Fuchs offers in place of ``$\psi$-ontology'' - i.e. in place of an ontology of quantum systems that exist in objective quantum states and evolve according to a well-defined objective dynamics.  QBism proposes instead an ontology of ``this and this and this ... every particular that is and every way of carving up every particular that is'' (p. 22) in which ``the real world ... is taken for granted'' (p. 7).  It is shown that this common-sense ontology fails to make contact with the implicit ontology of Fuchs' formalism, an ontology of quantum systems characterized by well-defined finite values of $d$, a quantity that Fuchs takes to be observer independent and hence to characterize a quantum system as a matter of objective fact: ``Dimension is something a body holds all by itself, regardless of what an agent thinks of it'' (p. 23).  This ontological disconnect raises the question of what Fuchs means by ``positing quantum systems as `real existences' external to the agent'' (p. 15), and suggests that Fuchs' ``positing'' of systems may not be common-sensical at all, but rather a fundamental and conceptually problematic axiom as it is in quantum Darwinism.  It is then shown that, while Fuchs dismisses decoherence theory and the general project of explaining the ``emergence'' of classicality \cite{zurek:98rough, zurek:03rev, joos-zeh:03, schloss:04, schloss:07} as ``unnecessary (and actually obstructive)'' (p. 24; parentheses in original), QBist and quantum Darwinist observers face a common problem of \textit{identifying} the quantum systems with which they are interacting, and to which, \textit{eo ipso}, the terms of their quantum-mechanical formalism refer.  Under both approaches they are forced into a common solution - basically Bohr's solution - of assuming that quantum systems can be identified by their classical properties \cite{bohr:28}.  This solution distorts the quantum formalism by forcing terms referring to quantum systems to function both as logically proper names and as definite descriptions, rendering their semantics incoherent and suggesting that diachronic term reference poses a deeper problem for quantum foundations that is generally recognized.   

The third and final stage examines the role played by observers in QBism.  QBist observers employ quantum mechanics as a ``users' manual'' (p 8) to cope with the uncertainty of the world, but by being autonomous, appear also to be the source of this uncertainty.  One might expect, therefore, for QBism to view the observer as a primary object of theoretical investigation.  Fuchs lampoons this suggestion: ``would one ever imagine that the notion of an agent, the user of the theory, could be derived out of its conceptual apparatus?'' (p. 8).  QBism nonetheless provides an implicit ``theory of the observer'' as an entity that is both an autonomous agent that uses quantum mechanics to update beliefs and an observable ``real-world'' quantum system with fixed $d$.  As will be shown, all quantum systems can be regarded as observers in QBism; hence the measurement environment itself can be regarded as an observer as it is in decoherence theory.  Characterizing the environment as an observer yields a QBist analog of the ``environment as witness,'' a ``super-observer'' that continuously monitors and updates its beliefs about all other systems.  It is shown that including the environment as a witness in the description of measurement places an intolerable strain on the QBist ontology, one that can be resolved only by merging the observed external system and the surrounding environment into a single entity to which the observer's beliefs refer.  The paper concludes by suggesting that this ontological predicament is not a specific side-effect of QBism, but rather a general consequence of quantum theory itself, one that appears to be resolvable only by rejecting the assumption of objectively separable systems to which the terms of the theory diachronically refer.

\section{QBist metaphysics}

As noted, quantum information for the QBist is about ``the consequences (for \textit{me}) of \textit{my} actions upon the physical system.''  The individual-specific Bayesian view of probability adds the ``for \textit{me}'' to the traditional instrumentalist and nomological-deductivist position that predictive power is all that can be expected of science, producing a view in which even probabistic predictions are purely personal: ``QBism finds its happiest spot in an unflinching combination of `subjective probability' with `objective indeterminism' '' (p. 8, fn. 14).  Such positions have always been resisted from more realist, or possibly just more curious quarters with a demand to know \textit{what} caused the ``consequences'' and \textit{how} it caused them.  Fuchs' fierce rejection of solipsism (p. 6, Fig. 1 and p. 19, Fig. 5) is borne out by his willingness to address these questions: Fuchs is a personalist about information, but not a personalist (i.e. not a solipsist) about metaphysics or ontology.  Unlike a solipsist, Fuchs insists that science is possible, even if is an intensely personal science; unlike a traditional instrumentalist, Fuchs also insists that science provides explanations, even though they cannot be reductive.  This section focuses on Fuchs' account of ``how?'' the consequences with which QBist agents must cope come about; the ``what?'' question is considered in the next section.

Fuchs is very clear on how consequences \textit{do not} come about: they do not come about by the linear evolution of the state vectors of quantum systems: ``QUANTUM STATES DO NOT EXIST'' and hence cannot evolve linearly or otherwise.  How consequences \textit{do} come about is explained mainly by an illustration (p. 6, Fig. 1) showing an observer - described as an ``agent'' - interacting with a quantum system.  The agent interacts with the system via a measurement device, which is depicted as a prosthetic hand ``to make it clear that (it) should be considered an integral part of the agent'' (p. 6, Fig. 1 caption).  The interaction is purely local: the measurement device makes contact with the system being measured.  Interaction requires two steps.  First, the agent acts on the system; this action is represented formally by a POVM, a generalized observable consisting of a set of operators $\lbrace E_{i} \rbrace$.  Second, the agent ``experiences'' a consequence, which is one of the operators $E_{k}$: ``the consequence = an experience, $E_{k}$'' (Fig. 1).  The \textit{selection} of $E_{k}$ from among the $\lbrace E_{i} \rbrace$ is the reaction of the quantum system to the agent's action $\lbrace E_{i} \rbrace$.  This action and reaction constitute the measurement: ``For a Quantum Bayesian, the only physical process in a quantum measurement is what was previously seen as `the selection step' - i.e., the agent's action on the external world and its unpredictable consequence for her'' (\cite{fuchs:11}, p. 12).  The quantum state $|\psi \rangle$ describing the system ``makes no appearance but in the agent's head; for it captures his degrees of belief concerning the consequences of his actions'' (p. 6, Fig. 1 caption).

Four aspects of this depiction of measurement interactions are worth noting.  First, what is depicted in Fig. 1 at least appears to be a \textit{physical interaction}.  Fuchs describes it as such in the passage quoted above: ``the only \textit{physical process} in a quantum measurement is what was previously seen as `the selection step''' (italics added).  If the measurement apparatus is ``an integral part of the agent'' and the measurement interaction is a physical processes, then the Bayesian updating of the agent's belief as a result of the measurement - the action of the selected $E_{k}$ on the belief $| \psi \rangle$ to produce a particular observationally-revised belief $E_{k}(|\psi \rangle)$ - must itself be a physical event.  This rules out one commonplace view, that ``there is no relevant change in \textit{anything} physical when he does so (i.e. when an agent updates a belief); the only changes are internal to the agent
ascribing the state'' (\cite{timpson:08}, p. 585, emphasis in original).  If the ``internal'' changes that implement belief updating are physical changes implemented by the selected POVM component $E_{k}$, as they must be if information is regarded as physically encoded \cite{landauer:99}, the argument that QBism dispenses with ``collapse'' simply by regarding quantum states subjectively becomes problematic.  The concern of Palge and Konrad \cite{palge:08} that Fuchs' formalism requires an infinite collection of ``readjustment'' operators arranged in a look-up table to compensate for the initial state of the apparatus is also elevated from a mathematical worry to a serious implementation issue.  

Second, while the events composing the measurement interaction are traditionally viewed from the perspective of the agent, they can also be viewed from the perspective of the system being measured.  Viewing the agent as a physical system and using for a moment the traditional language of objective quantum states to describe the agent's states of ``belief,'' the agent is initially in $|\psi \rangle$, and is then acted upon by the system in a way that results in the agent being in $E_{k}(|\psi \rangle)$.  Viewed formally, this \textit{selection by an external quantum system} of a particular state $E_{k}(|\psi \rangle)$ of an agent who interacts with that system is a process familiar from decoherence theory: einselection, the selection by the environment of a particular quantum state of any system interacting with that environment \cite{zurek:98rough, zurek:03rev}.  Since this change of perspective on the situation employs the idea of a quantum state as an objective physical state ($|\psi \rangle$) of a physical system (the agent), it is completely opposed to the QBist view, but it illustrates the formal relationship between QBism and decoherence that Fuchs derives by other means \cite{fuchs:11}.

The third aspect worth noting follows from the second: given that in QBism $|\psi \rangle$ is \textit{not} an objective quantum state of the agent - ``quantum states do not exist'' - what kind of state is it?  What does Fuchs \textit{mean} by saying that $|\psi \rangle$ represents ``degrees of belief'' that are ``in the agent's head'' and how does the ``experience'' $E_{k}$ act on these ``degrees of belief''?  Quantum Bayesians, like ordinary Bayesians, cash out ``degrees of belief'' in terms of an agent's overt \textit{behavior}: choices made, decisions taken, bets placed and the like.  Behaviors are physical events, as are the sensory processes - photons impacting retinal cells, for example - that are required in order for measurement outcomes to be experienced.  That such physical events cause and are caused by transitions between ``degrees of belief'' suggests strongly that such transitions and hence ``degrees of belief'' themselves have some physical implementation within the physical systems constituting observers.  If they are not implemented by objective quantum states, how are they implemented?  This is clearly a question that Fuchs does not want to answer, as it calls for a theory of the agent of precisely the kind that Fuchs rejects.  The need for such a theory becomes clear, however, as soon as one imagines an ``observer'' that is not a common-sense cognitive agent.  Fuchs quotes with approval John Bell's lampoon of the idea that measurements are the distinct perogative of human beings or even living systems (p. 1), and later proposes that the observer - system relationship holds for ``every two parts of the world'' (p. 27), in which case \textit{every} physical interaction counts as ``measurement,'' just as it does in decoherence theory.  In place of the ordinary physical interaction between $A$ and $B$ by which ubiquitous measurement is implemented in decoherence theory, however, QBism substitutes the action by $A$ on $B$ with a POVM and the selection by $B$, in reaction, of a particular POVM component as $A$'s experience.  Hence \textit{all} physical systems, even elementary particles, must be capable of experience, and interpretable as having ``degrees of belief'' that these experiences update.  ``Does this mean even `elementary' physical events just after the big bang must make use of \textit{concepts} that, to the reductionist mind, must be 15 billion years removed down the evolutionary chain?  You bet it does'' (p. 21, fn. 37; emphasis in original).  Fuchs hints that the causal effects of such ``concepts'' and of ``degrees of belief'' in general may be ``example(s) of an interaction between two nonreductionist realms'' in which ``Each
realm influences the other as its turn comes'' (p. 21, fn. 37) as in classic Cartesian dualist interactionism.  If interactions between physical systems are to be regarded as \textit{physical} interactions and accounted for by \textit{physical} theory, however, ``concepts,'' ``degrees of belief,'' POVM components and ``experiences'' must all have physical implementations.

The fourth aspect of Fuchs' depiction of measurement that is worth noting is that measurement, for the QBist, does not involve quantum information exchange.  In QBism, quantum information is personal information ``for \textit{me}'' that is inaccessible to other agents, as Fuchs makes clear in his extended analysis of Wigner's friend (p. 6-7; see also below).  This distinguishes QBism from alternative, non-personalist instrumentalist approaches such as \cite{bub:04} as well as more traditional approaches in which (objective) classical information is (objective) quantum information that has survived decoherence \cite{zurek:03rev, schloss:07, griffiths:07}.  Hence while Fuchs comments briefly on the reasonableness of viewing quantum teleportation in the usual way, i.e. saying that ``it transfers \textit{quantum information} from Alice's site to Bob's'' (p. 3; emphasis in original), this view is not consistent with QBism.  If quantum information is non-sharable personal information, Bob may acquire it by acting on a quantum system with a POVM, but it cannot be transferred from one agent to another.  In the case of quantum teleportation, the quantum system on which Bob acts with a POVM, and which selects an ``experienced'' POVM component in return, is the system that implements what would ordinarily be called the communication channel from Alice, but since quantum information cannot be communicated, this system does not \textit{function} as a channel in the scenario.  Bob's interaction is therefore a \textit{local interaction with the channel}, not a distant, channel-mediated interaction with Alice.  How Bob's interaction with the channel selects a POVM component that acts on his prior beliefs about Alice - or even his prior beliefs about the channel - is not described by QBism; POVMs are usually defined as acting on quantum states (e.g. \cite{nielsen-chaung:00}, Ch. 2), and quantum states do not exist.  However, by making quantum information personal and thereby ruling out quantum information exchange between systems, Fuchs appears to rule out any \textit{physical} selection mechanism, as any such mechanism could be described in terms of quantum information exchange, as indeed they are described in more traditional approaches to measurement.  Hence as with the question of how ``experiences'' are implemented physically discussed above, how ``selection'' of an experience is implemented remains to be explained within QBism.

In the absence of a well-defined physical mechanism for POVM component selection, Fuchs lapses into language that appears, at least, to be entirely metaphorical: ``quantum measurements are \textit{moments of creation}'' (p. 14; emphasis in original); their consequences are ``unique creation(s)'' (p. 6, Fig. 1 caption) to which quantum systems ``give birth'' (p. 27).  Consistent with this view of measurement interactions as involving ``creation'' and ``birth,'' \textit{all} quantum systems are characterized as ``agents'' that, within the constraints of their finite values of $d$, are \textit{autonomous}: ``the universe ... should be thought of as a thriving community of marriageable, but otherwise autonomous entities'' (p. 14).  This sense of autonomy is stronger than the intrinsic stochasticity of e.g. \cite{mermin:98} or \cite{griffiths:11}: autonomy here indicates objective, in-principle indeterminism with a corresponding in-principle absence of external control.  As Timpson puts it, ``the really crucial conception is that what (singular) event will occur when a system and a measuring device interact is not determined by anything, not even probabilistically'' (\cite{timpson:08}, p. 596).  Fuchs' sense of objective indeterminsm goes beyond even that of Cartwright \cite{cartwright:99}: ``where things differ in our case (i.e. QBism) is that we are imagining that at the fundamental level there are \textit{no} situations, however specialised, in which we will obtain lawlike behaviour'' (\cite{timpson:08}, p. 597).  Fuchs uses the term ``interiority'' (p. 20) to emphasize the inaccessibility of an individual-specific quantum state (i.e. a belief) to inspection by external agents or manipulation, even probabilistically, by external forces.  Such autonomy is the very antithesis of the deterministic, unitary time evolution described by the Schr\"odinger equation, and applied to the universe as a whole by Everett \cite{everett:57} and all advocates of minimal quantum mechanics since (e.g. \cite{tegmark:10, schloss:06}; for broader discussion see \cite{schloss:07, landsman:07, wallace:08}). 

Universal autonomy, applying to all systems at all scales, would imply that \textit{no} dynamics are well defined either within or between specified systems.  Fuchs is committed to the autonomy of human observers, quoting with approval a statement by Hans Primus that such autonomy is a prerequisite of science (p. 18).  Given this committment for human observers, the Conway-Kochen ``free will'' theorem requires it to be extended to quantum systems in general \cite{conway-kochen:06, conway-kochen:09}.  Fuchs frames universal autonomy formally as a rejection of the EPR ``criterion of reality'' that a probability-one prediction by a correct dynamical theory \textit{requires} that an event be observed if anyone bothers to look for it (p. 18, esp. fn. 29), but his gloss on this result strips away any doubt that it is just an epistemological claim: ``At the instigation of a quantum measurement, something new comes into the world that was not there before; and that is about as clear an instance of \textit{creation} as one can imagine.  Sometimes one will have no strong beliefs for what will result from the creation ... and sometimes one will have very strong beliefs ... but a free creation of nature it remains'' (p. 19; emphasis in original).  How such ``free creations'' relate to the apparently physical processes ilustrated in Fig. 1 or to the action on the initial belief $| \psi \rangle$ by the selected POVM component $E_{k}$ to produce a revised belief $E_{k}(|\psi \rangle)$ is never discussed.

With ``free creation'' replacing well-defined dynamics, there can be no idea of well-defined physical mechanisms by which either the system or the observer evolve that can be discovered by empirical investigation.  Physical ``laws'' can be discovered, but these are normative statements that guide descision making, not descriptions of objective facts.  The Born rule, for example, is ``normative ... like the Biblical Ten Commandments''; it ``guides, `Gamble in such a way that all your probabilities mesh through me' '' and agents are ``free to ignore the advice'' if they choose to (p. 8, Figure 2 caption).  Hence the fundamental questions raised by Fuchs' illustration of measurement, \textit{how} does the agent's action with the POVM affect the system, and \textit{how} does the system ``select'' a particular POVM component as the agent's experience, not only are not but cannot be answered within QBism.  Fuchs redefines the goals of physics away from answering such questions, first by rejecting any search for the ``building blocks'' of the world: ``Physics ... is not about identifying the bricks with which nature is made, but about identifying what is \textit{common to} the largest range of phenomena it can get its hands on'' (p. 22; emphasis in original).  Newton's law of universal gravitation is advanced as an example of such a ``non-reductionist'' commonality (though how Newton's quite deterministic law would fare in a universe of autonomous gravitational agents is not discussed).  Many would regard descriptions of ``underlying'' physical mechanisms - gauge boson exchange, for example - as exemplary physical commonalities, but mechanisms by definition require not just building blocks, but also the notions that buildings blocks have both individual and collective states, and that these states evolve in well-defined ways.  Fuchs rejects all of these.  He congratulates Bohr for refusing to give a mechanistic explanation for atomic spectra (p. 25), alluding (one assumes) to Bohr's admission that ``there obviously can be no question of a mechanical foundation'' for his quantized model of electron orbitals (\cite{bohr:13}, p. 15).  By refusing himself to provide any mechanistic account of the selection of POVM components by physical systems or of the updating of beliefs (i.e. quantum states) by observers, Fuchs at least appears to place measurement, and hence physical interactions in general, outside the realm of physical investigation.   

In the place of common building blocks or physical mechanisms, Fuchs proposes the dimension $d$ as the commonality, the ``previously unnoticed capacity inherent in all matter'' (p. 23) that quantum mechanics has discovered, and that quantum theory is about.  QBism makes the bold claim that the dimension $d$ of any quantum system is by definition finite; that the infinite dimensional Hilbert spaces commonly attributed to quantum systems are just ``a useful artifice when a problem can be economically handled with a differential equation'' (p. 25).  Dimension appears, for Fuchs, to measure the ability of a quantum system to behave autonomously, to ``create.''  He characterizes $d$ as a measure of ``the fuel upon which quantum computation runs,'' ``the raw irritability of a quantum system to being eavesdropped upon'' (both p. 23), ``oompf'' (p. 25) or ``zing'' (\cite{fuchs:02}, p. 9).  Though Fuchs points enthusiastically toward the future, the kinds of physics one can do with autonomous agents characterized by fixed, finite values of $d$ remain unclear.  Approaching this question requires understanding what these autonomous agents \textit{are}, i.e. understanding ``\textit{just what after all is the world made of}'' (p. 9; emphasis in original) for a QBist.

\section{QBist ontology}

As mentioned above, Fuchs assumes an ontology in which ``the real world ... is taken for granted'' and adds to this ``common sense'' ontology the notion of a quantum system with a well-defined, observer-independent dimension $d$.  Understanding ``what'' causes the consequences that QBist observers observe requires understanding how being an ordinary object, a ``this and this and this'' relates to being a quantum system with fixed $d$.  This section shows both that this relationship is not as straightforward as Fuchs appears to think, and that QBism shares this fundamental ontological non-straightforwardness with quantum Darwinism \cite{zurek:98rough, zurek:03rev, zurek:04, zurek:05, zurek:06, zurek:07grand, zurek:09rev}, perhaps the most distant from QBism of all major approaches to quantum foundations.  The fundamental issue underlying this non-straightforwardness lies deep in the formalism of quantum mechanics: it is the question of what is implied operationally by the assignment of a name - e.g. $S$ or $Alice$ - to a quantum system.  This section makes precise the ontological problem of specifying the referent of a formal expression that names a quantum system as it arises in QBism, and compares the QBist version of this reference problem to the version that arises in quantum Darwinism.  It shows that QBism cannot have both system names that refer diachronically in a ``common sense'' way and fixed values for system dimension, and raises the question of how QBism is to cope with the common scientific practice of defining systems \textit{ad hoc}.  The next section explores these issues from the perspective of a QBist observer, and shows that they can be resolved only by viewing every QBist observer as interacting with exactly one system: the entire rest of that observer's universe. 

What exactly is being assumed by QBism when ``the real world ... is taken for granted''?  Fuchs is clearly serious on this point; he stresses that quantum mechanics is a tool for making practical choices in the real world characterized by real uncertainty: ``In my case, it is a world in which \textit{I} am forced to be uncertain about the consequences of most of \textit{my} actions; and in your case, it is a world in which \textit{you} are forced to be uncertain about the consequences of most of \textit{your} actions'' (p. 8; see also p. 20, fn. 33).  Following Einstein and many others, Fuchs explicitly ``takes for granted'' both separability and local causation: ``what is it that A and B are spatially distant things but that they are causally independent?'' (p. 15).  Consistent with the idea that quantum states have a particular space-time location - in the head of the agent whose beliefs they characterize (p. 5) - EPR pairs and entanglement in general do not objectively exist in QBism; indeed, preserving Einsteinian intuitions of locality and separability are two of the major motivations behind QBism \cite{fuchs:02}.  If an agent assumes two quantum systems are separable, they are: ``If the left-hand system can manipulate the right-hand system, \textit{even when by assumption it cannot}, then who is to say that the right-hand system cannot manipulate the agent himself?  It would be a wackier world than even the one QBism entertains'' (p. 18; emphasis in original).  Hence ``correct dynamical theories'' cannot \textit{require} external systems to be in particular states, as the EPR ``criterion of reality'' would suggest, but assumptions by agents can \textit{require} that systems be separable, ruling out entanglement as a real, agent-independent phenomenon.  The assumptions of separability and locality across the board mean that for Fuchs, not just apparatus (as \cite{bohr:28} insisted) but \textit{all systems} can be considered to be ordinary classical ``objects''; what quantum mechanics captures is any observer's in-principle uncertainty about what any one of these objects, being an autonomous (quantum) agent, will do next.  Hence what is ``taken for granted'' by QBists is what the objects making up the ``real world'' \textit{are}; what these objects might \textit{do} is, for the QBist, fundamentally uncertain.

Fuchs uses the classic paradox of Wigner's friend to illustrate the subjective nature of beliefs about, and hence quantum state assignments to, ordinary ``real world'' objects.  He introduces the paradox in the usual way: ``It starts off with the friend and Wigner having a conversation: Suppose they both agree that some quantum state $|\psi \rangle$ captures their mutual beliefs about the quantum system'' (p. 6), and goes on to discuss how Wigner's and his friend's individual quantum state assignments (i.e. beliefs) diverge after the friend performs a measurement unobserved by Wigner.  What is interesting here, and what remains undiscussed by Fuchs (or other commentators on Wigner's friend, e.g. \cite{schloss:07}, p. 364-365 or \cite{timpson:08}, p. 584-5 and 593), is that (1) Wigner and his friend are able to identify a \textit{single} quantum system as a mutual ``system of interest'' to which their beliefs refer, (2) that the friend is able, after the conversation with Wigner, to carry out a measurement on \textit{that very same} quantum system (a perdurantist would say, \textit{part of} the very same system), and (3) that after the friend's measurement, the friend and Wigner can potentially disagree about - i.e have different beliefs that refer to - \textit{the same system} that they previously mutually identified.  How does this work: how do Wigner and his friend both \textit{identify} the system of interest, and how do they \textit{re-identify} it (or if they are perdurantists, its later temporal parts) later?  It appears that, as with separability, if Wigner and his friend \textit{assume} that they are referring to the same system, then they \textit{are} referring to the same system; similarly if they assume they have re-identified the same system, then they have re-identified it, as a matter of objective, physical fact.  These assumptions are made very clear in the formalism: Wigner treats his friend as a second quantum system to which he assigns a quantum state (i.e. a belief) $\rho$, he assumes a unitary time evolution $U$ for the combined $friend \otimes system$, and computes an expected joint state $U( \rho \otimes |\psi \rangle \langle \psi |)U^{\dagger}$ (p. 7, Fuchs' notation).  This formal representation explicitly assumes that $U( \rho \otimes |\psi \rangle \langle \psi |)U^{\dagger}$ refers to $friend \otimes system$, i.e. that $U$, being a unitary operator defined on the Hilbert space of $friend \otimes system$, does nothing that would alter the \textit{identity} over time of $friend \otimes system$ or its Hilbert space.  It also implicitly assumes that whoever calculated it is still able to identify $U( \rho \otimes |\psi \rangle \langle \psi |)U^{\dagger}$ as a state of $friend \otimes system$, i.e. that it is clear to observers that the time evolution has done nothing to alter the identity of $friend \otimes system$.  These formal representations would clearly be incorrect if Wigner was mistaken about \textit{which system} his friend had identified and measured, or if he was mistaken in his re-identification of his friend.  

Before considering the question of \textit{how} observers determine that a system seen now is the same thing as a system seen earlier within QBism, let us examine this question from the more traditional perspective of ``$\psi$-ontology'' and objective quantum states.  Within this framework, a Hilbert space for a system $S$ is specified by specifying a set of basis vectors $\lbrace |s_{i}\rangle \rbrace$; the dimension of the Hilbert space is the number of these basis vectors.  Each basis vector is taken to represent a physical degree of freedom that the system has as a matter of objective fact.  If the system has many parts that can move or otherwise behave independently, then the individual ways that these parts can move or otherwise behave are counted among the degrees of freedom.  The state of the system is then represented as a function of these degrees of freedom by an expression such as ``$|S\rangle = \sum_{i} \lambda_{i} |s_{i}\rangle$''; this state is an objective characteristic of the system observable, as Pauli says, ``by objective registering apparatus, the results of which are objectively available for anyone's inspection'' (p. 7, quotation by Fuchs).  This expression \textit{formally} specifies the identity of $S$; to point and say ``\textit{this} is $S$'' is just to say that ``this'' has the specified physical degrees of freedom and no others, and that its objectively existing state has the form $\sum_{i} \lambda_{i} |s_{i}\rangle$ for some set of complex numbers $\lambda_{i}$.  

In practice, in the laboratory, no one examines systems such as particle detectors or laptop computers to determine what physical degrees of freedom they have, and no one checks that the laptop computer, for example, on their desk today has exactly the same set of physical degrees of freedom that the laptop computer on their desk yesterday had.  The identity over time of these systems is $assumed$, just as Wigner $assumes$ that the person that he sees when he walks back into the laboratory is indeed his friend.  Formally, this is an assumption that the propagation of the system over time is unitary; in practice, it is the assumption that the parts composing the system and the physical degrees of freedom of those parts have not changed.  Evidence that either the parts or their physical degrees of freedom \textit{have} changed can lead questions about the system's identity and demands for an explanation.  However, obtaining such evidence requires first identifying the system, at least tentatively: to determine that the laptop on your desk is not the same one that was there yesterday, one first has to identify it as \textit{something}, i.e. as not part of the general background of the world, and then identify it as a \textit{laptop}.  Employing a measurement device to obtain information about some distant, microscopic, or otherwise not directly observable system similarly requires first identifying the measurement device: a pointer value of ``5,'' for example, means nothing on its own.  Traditional accounts of measurement do not explicitly address this issue of system identification; they assume that systems are specified by specifying a list of objective degrees of freedom, and go on from there. 

Fuchs carefully avoids any discussion of physical degrees of freedom throughout ``QBism: The perimeter of quantum Bayesianism''; indeed his discussion of ``non-reductionist'' goals for physics suggests that he views investigations of the parts of things and how they might move or otherwise behave to be outside the proper purview of experimental science.  Hilbert spaces are not objective collections of physical degrees of freedom in QBism, but rather calculation tools; it is only ``dimension'' that is objective.  Hence in QBism, physical systems cannot be identified by examining their parts or the physical degrees of freedom of such parts.  One cannot check that $|S\rangle = \sum_{i} \lambda_{i} |s_{i}\rangle$ as a matter of fact, because $|S\rangle = \sum_{i} \lambda_{i} |s_{i}\rangle$ is not a matter of fact, it is only a subjective, personal \textit{belief}.  The ``real world,'' however, is nonetheless real, and friends, laptops and laboratory apparatus have to be identified.  Within QBism, all of these things are systems, and systems are autonomous agents that are probed with POVMs and respond by selecting POVM components.  Hence system identification can only be accomplished, within QBism, by probing systems with POVMs.  Because POVMs are formally defined over particular Hilbert spaces, each particular system within the ``real world'' must have a particular POVM that identifies it and only it.  The assumption that ``the real world ... is taken for granted'' must, therefore, be supplemented by an additional assumption: that observers have at their disposal POVMs capable of specifically identifying the systems that populate the world \cite{fields:12}.

How this supplementary assumption works in practice can been examined in the case of Wigner's friend.  In order to identify the system of interest and his friend, Wigner must be equipped with POVMs - call them ``$\lbrace S_{i} \rbrace$'' and ``$\lbrace F_{i} \rbrace$'' respectively - specific to each of them.  Suppose that when Wigner acts on the system and his friend with these identifying POVMs at $t$, they select the POVM components $S_{k}$ and $F_{k}$ in reaction.  At some later time $t + \Delta t$, Wigner employs these POVMs again, and the same POVM components are again selected in reaction.  To infer that the systems probed at $t + \Delta t$ were the same systems that were probed at $t$, Wigner must assume that the POVMs $\lbrace S_{i} \rbrace$ and $\lbrace F_{i} \rbrace$ yield unambiguous identifications, i.e. that the components $S_{k}$ and $F_{k}$ can only be selected by the system of interest and his friend, respectively.  This assumption can clearly take two forms: it can be assumed either that $\lbrace S_{i} \rbrace$ and $\lbrace F_{i} \rbrace$ are \textit{in fact} specific to the system and the friend, respectively, or that these POVMs are defined over a larger Hilbert space that encompasses these objects plus others, but that they nonetheless yield the components $S_{k}$ and $F_{k}$ only when acting on the system and the friend, respectively.  As Hilbert spaces are not objective in QBism, there is no \textit{physical} difference between these two assumptions.  For simplicity, therefore, it can be assumed that $\lbrace S_{i} \rbrace$ and $\lbrace F_{i} \rbrace$ are individual-specific detectors with only two significant components: $S_{k} = S_{1} = 1$ is selected always and only by the system of interest (similarly $F_{k} = F_{1} = 1$ by his friend) and $S_{0} = 0$ is selected always and only by systems other than the system of interest (similarly $F_{0} = 0$ by systems other than his friend).  These POVMs then behave semantically as diachronic rigid designators, i.e. logically proper names.  They must commute with each other and with all other observables (at least all other observables applicable to the system or the friend), i.e. they must behave as \textit{classical} diachronic rigid designators.  This requirement is independent of what kind of system is being referred to; it applies equally to ``hunks of matter'' and to ``higher-order'' entities identified by structural or functional criteria (e.g. \cite{wallace:05}).  Because QBism has no concept of an objective quantum state and hence no concept of a projection defined as a physical process acting on objective quantum states, it has no natural way of imposing an objective coarse-graining that would permit a many-to-one mapping from quantum systems to observations other than equality of selected POVM components.  Hence QBism has no concept of a ``framework'' within which multiple systems might yield indistinguishable observations (e.g. as in the consistent histories approach \cite{griffiths:11}) other than that provided by POVM component selection.

It is not only Wigner who must assume that Wigner has available POVMs that identify individual quantum systems diachronically; to develop his critique of ``$\psi$-ontology'' Fuchs must assume this, too.  For if either or both of Wigner's diachronic identifications are wrong, his computed belief $U( \rho \otimes |\psi \rangle \langle \psi |)U^{\dagger}$ is, from his perspective, no longer about $friend \otimes system$, but about whatever Wigner has mistakenly identified as $friend \otimes system$.  In this case, the paradox of Wigner's friend collapses; nothing of interest follows from Wigner and his friend having inconsistent beliefs that refer, in fact, to different external systems.  The possibility of conflicting beliefs about a single quantum system tells against the objectivity of quantum states only if the conflicting beliefs are assumed to refer to the \textit{correct} systems.  QBism thus not only requires that the systems composing the world can be ``taken for granted'' and that POVMs that specifically identify these systems exist; it  also requires that observers employ these system-identifying POVMs correctly - it requires that the ability of observers to \textit{name things} consistently over time be taken for granted.  This assumption that system identification and hence term reference works in a straightforward way is an unnoticed but natural correlate of the assumptions of separability and locality.

The first thing to note about the diachronic system-identifying POVMs tacitly assumed by QBism is that they are required not just formally but operationally: they must be implemented by the local observer - system interactions of the kind illustrated in Fuchs' Fig. 1.  The constitutions of Wigner and his friend, for example, must be such that when Wigner executes the action $\lbrace F_{0}, F_{1} \rbrace$ on his friend, his friend selects $F_{1}$ as Wigner's experience in reaction; otherwise Wigner cannot identify his friend.  The strength of this requirement becomes clear when one consider the usual channel-mediated Alice - Bob interaction, e.g. quantum teleportation.  In this case, Bob's local action with an Alice-identifying POVM $\lbrace A_{0}, A_{1} \rbrace$ on the \textit{channel} must physically select $A_{1}$ as Bob's experience in reaction if Bob is to identify the channel as the channel \textit{from Alice}.  As noted above, QBism offers no account of how this implementation of POVM component selection works.  Any \textit{physical} implementation, however, introduces the physical possibility of error; if an action by Wigner on some other system $X$ with $\lbrace F_{0}, F_{1} \rbrace$ happens to result in a selection of $F_{1}$, Wigner will mistakenly identify $X$ as his friend, a situation not unknown in real life (e.g. \cite{sacks:85}).  Hence if it is to admit any physical interpretation, the most that QBism can actually assume is that system-identifying POVMs implement diachronic designators that are approximately rigid, that agents using such identifiers get system identification right most of the time, and hence that Wigner's friend scenarios remain paradoxical for realist interpretations of quantum states most of the time.  

The second thing to note is that interactions with the operational characteristics of $\lbrace F_{0}, F_{1} \rbrace$ are also tacitly assumed by other interpretative approaches to quantum mechanics, including in particular quantum Darwinism.  In the ``environment as witness'' framework, the locus of this assumption is the operational definition of ``objectivity'':

\begin{quotation}
``A property of a physical system is \textit{objective} when it is:
\begin{list}{\leftmargin=2em}
\item
1. simultaneously accessible to many observers,
\item
2. who are able to find out what it is without prior knowledge about the system of interest, and 
\item
3. who can arrive at a consensus about it without prior agreement."
\end{list}
\begin{flushright}
(p. 1 of \cite{zurek:04}; p. 3 of \cite{zurek:05})
\end{flushright}
\end{quotation}
 
This definition, like Fuchs' recounting of Wigner's friend, assumes that the observers in question can arrive at a consensus that they are observing the same system and maintain that consensus through the process of determining the system's properties.  Hence it requires that the observers have an operational method of identifying the system at the start of their investigations and re-identifying it at the end.  This operational re-identification method must commute with all other observables, and hence must be effectively classical \cite{fields:10, fields:11} just as it does in QBism.  Quantum Darwinism is realist about quantum states and realist about decoherence, while QBism is instrumentalist about both; the common assumption of diachronic (approximately) rigid designators by both approaches shows that this assumption is independent of whether one is a realist or an instrumentalist about quantum states.

The third thing to note about QBism's requirement for individual-specific diachronic identifiers is that it conflicts with QBism's basic postulate that each quantum system has a well-defined, finite, observer-independent dimension $d$.  As noted above, Fuchs introduces $d$ as the Hilbert-space dimension of a system (p. 12; see also p. 19 ff and \cite{fuchs:02} p. 51) and explicitly argues that $d$ as a Hilbert space dimension is always finite (p. 25), but throughout refers to $d$ simply as ``dimension'' to emphasize that Hilbert spaces are not objective entities in QBism but only tools for performing calculations.  The capacity of a system for autonomous behavior that $d$ measures is, however, an objective fact about the world; it is the ``previously unnoticed capacity inherent in all matter'' that Fuchs claims quantum mechanics has discovered (p. 23).  The conflict between the need for individual-specific diachronic identifiers and fixed, objective values of $d$ appears as soon as one applies both requirements to a quantum system that continues as a ``real existence'' through time.  Suppose that $X$ is such a system, and that an agent $A$ has a POVM $\lbrace X_{0}, X_{1} \rbrace$ that uniquely and accurately identifies $X$ at least most of the time.  Because $d$ \textit{objectively} measures the capacity for autonomy and the dimension $d_{X}$ of $X$ is fixed, any beliefs that $A$ may have, however they have been obtained, that include attributions of more or less capacity for autonomy to $X$ at different times are false.  This is clearly paradoxical if the individual-specific diachronic identifiers required by the theory function at all like proper names in a natural language, i.e. if they continue to refer despite significant structural and featural changes to their referents.  Consider, for example, $A$'s beliefs concerning her children: the capacity of children for autonomous behavior increases with time on any plausible reading of ``autonomous,'' yet proper names in natural languages continue to refer to them despite their increasing $d$, or in the traditional, non-QBist view, the increasing numbers of objective physical degrees of freedom in their objective Hilbert spaces. While non-QBist quantum mechanics can fall back on a \textit{definition} of system identity based on the presence or absence of physical degrees of freedom, however, QBism does not have this option; QBism can only define diachronic system identity using individual-specific POVMs.  If these POVMs do not continue to refer despite significant featural changes to their referents, which they will not if the systems to which they refer do not have fixed $d$, it is not clear how they can support the communication or consensus needed to support the diachronic identification of a system $X$ by multiple agents who have mutually observationally inaccessible experiences $X_{1}$ or $X_{0}$.  Without this ability, however, system-identifying POVMs cannot do their job: they cannot identify systems as ``real existences'' in the real world. 

Significant changes in the properties of systems named by natural-language identifiers are often handled, in practice, by ``everyone knows'' conventions about the operational significance of the changes: ``everyone knows'' that the behavioral capacities of the LHC have changed over the lifetime of the term ``LHC'' and ``everyone knows'' that they must take these changes into account when acting on beliefs about the LHC.  This work-around hides the problem of diachronic reference, however, rather than solving it.  If $d$ is fixed, the behavioral capacities of systems never change; what is called ``behavioral change'' in ordinary language - for example, the change from not being able to accelerate protons to several TeV to being able to so accelerate them - is the replacement of a system $X$ with dimension $d$ by a different system $X^{\prime}$ with dimension $d^{\prime}$.  In a QBist context, the problem of diachronic reference is solved only if all agents concerned with a given system $X$ update their system-identifying POVMs whenever $X$ is replaced \textit{in fact} by $X^{\prime}$.  Even setting aside the question of how agents are to \textit{discover} that $X$ has been replaced by $X^{\prime}$, how they update their system-identifying POVMs remains a mystery: an agent can attempt to act on any system at all with a POVM $\lbrace X^{\prime}_{0}, X^{\prime}_{1} \rbrace$, but that POVM does not identify $X^{\prime}$ unless $X^{\prime}$ reacts back on the agent to select $X^{\prime}_{1}$.  Hence it would appear that updating an identifier, in QBism, requires a \textit{physical} operation on $X^{\prime}$, a system that the agent attempting to perform the update cannot, by assumption, identify.

The problem of diachronic reference can be solved, within QBism, only by admitting that agents routinely get it wrong, or by dropping the requirement that systems have fixed $d$.  Either of these options raises, rather forcefully, the question, ``what is a \textit{system}, anyway?''  Taking the ``real world'' for granted is clearly insufficient for answering this question; just pointing to ``this and this and this'' does not work either unless one is prepared to \textit{abandon language}, including the mathematical language of the quantum-mechanical formalism, in its diachronic referential role.  Fuchs says very little about the nature of systems, except to imply that any bounded part of the world, any ``real existence'' or in traditional non-QBist language, any bounded collection of physical degrees of freedom counts as one: ``every way of carving up every particular that is'' (p. 22) counts as a system, and ``in every `hole' (every bounded region) there is an interiority not given by the rest of the universe and a common quality called dimension'' (p. 25; parentheses in original).  If systems are ``real existences,'' then the boundaries that separate them from the rest of the universe - from the ``environment'' in decoherence-theory terminology - must be ``real existences'' too: they must \textit{physically} keep dimension in and keep deterministic causal influences out, providing the (at least partial) isolation or ``interiority'' required for autonomy.  Hence the question, ``what is a system?'' becomes the question ``what is a system \textit{boundary}?'' in QBism.  

QBism is clearly not the only foundational approach that assumes physical system boundaries that play a causal role: system boundaries define the regions of Hilbert and typically physical space from which coherence is removed in decoherence theory.  Zurek explicitly recognizes their problematic status, noting that ``a compelling explanation of what the systems are ... would undoubtedly be most useful'' (\cite{zurek:98rough}, p. 1818) and in later papers postulating as ``axiom(o)'' of quantum mechanics that ``systems exist'' (\cite{zurek:03rev}, p. 746; \cite{zurek:07grand}, p. 3).  As in the case of system identification by observers, the emergence of a fundamental problem shared by foundational approaches as diametrically opposed as QBism and quantum Darwinism suggests a deeper underlying issue: it suggests that the assumption that systems and their boundaries are ``real existences,'' convenient as it is from a formal perspective, may be an error.  In the ordinary practice of science, after all, systems are routinely defined as needed to address particular problems; no one asks whether a particular boundary picking out some subset of objects that happen to be of interest is ``real.''  Before the advent of the ``environment as witness,'' decoherence was explained informally as a consequence of the observer \textit{not looking} at the environment (formally, tracing the environment out of the density matrix), and hence ``losing'' the coherence information that system-environment entanglement transferred to environmental degrees of freedom \cite{zeh:70, zeh:73, zurek:81, zurek:82, joos-zeh:85, zeh:06}.  Fuchs lampoons the idea of observer-dependent decoherence (24, fn. 45), but in doing so commits himself to the view that all observer-specified systems, whatever their purpose or shelf-life, are ``real existences'' with real boundaries capable of preserving autonomy, i.e. keeping not just the prying eyes of observers but the causal forces of the rest of the universe at bay.  How this view of systems and their $d$-preserving boundaries as ``real'' is to cope with the profusion of arbitrarily bounded and arbitrarily overlapping systems postulated \textit{ad hoc} by observers, or with ``every particular that is and every way of carving up every particular that is'' as defined by Nature herself, is not made clear.  As all QBist systems are agents and hence observers, examining this question requires better understanding the abilities that Fuchs attributes to observers.

\section{What is a QBist ``observer''?}

Unlike most recent approaches to quantum foundations, QBism does not attempt to ``remove the observer from the theory just as quickly as possible'' (p. 2) in order to obtain an observer-independent formalism.  QBist observers occupy center stage in the theory, placing bets, acting on quantum systems with POVMs, and experiencing unpredictable consequences.  QBist observers are not passive recipients of information, as they arguably are in quantum Darwinism, but active, autonomous \textit{agents}.  For the QBist, the quantum mechanical formalism provides a ``user's manual for decision-making agents immersed in a world of \textit{some} yet to be fully identified character'' (p.  23; emphasis in original) that contains normative rules for calculating the probabilities of outcomes, in particular the Born Rule (p. 8, Fig. 2).  Using this rule, an agent can calulate consistent sets of beliefs about quantum systems (p. 4) and hence avoid bad outcomes (p. 8).  The quantum mechanical formalism is universal: ``It is a users' manual that \textit{any} agent can pick up and use to help make wiser decisions in this world of inherent uncertainty'' (p. 8; emphasis in original).

Fuchs does not explicitly define ``agent'' and insists that ``the notion of agent ... cannot be and should not be derivable from the quantum formalism itself'' (p. 20) and more broadly that ``agency, for sure, is not a derivable concept'' (p. 27).  What counts in QBism as an agent must, therefore, be reconstructed from Fuchs' characterizations of what agents do and how they do it.  The first and most obvious question is, are all QBist agents quantum systems and hence ``real existences''?  The idea that observers are themselves quantum systems and can be treated theoretically as such dates back at least to \cite{vonNeumann:32} and is taken for granted in most discussions of quantum foundations.  Fuchs appears to accept this notion, but then reject it: ``It will be noticed that QBism has been quite generous in treating agents as physical objects when needed.  `I contemplate you as an agent when discussing your experience, but I contemplate you as a physical system before me when discussing my own' '' (p. 27).  The term ``contemplate'' suggests that Fuchs views the distinction between an ``agent'' and a ``physical system'' to be a \textit{semantic} distinction, a matter of interpretation.  The quoted passage is followed by one in which Fuchs acknowledges that all physical systems are autonomous, and hence agent-like in QBism.  As noted in Sect. 2, the universality of agency in QBism follows directly from Fuchs' postulate of universal ``objective indeterminism'' as well as from his assumption of agency for human observers via the Conway-Kochen theorem.  It appears, then, that Fuchs would be willing to grant that all agents are quantum systems, but only with the provisos that (1) all quantum systems are agents, and (2) ``agent'' and ``quantum system'', even if co-extensive, have different (unspecified) connotations within QBism.

If all quantum systems are agents, any measurement interaction can be viewed from either direction: a human observer monitoring the beam current in the LHC using a PC can equally be viewed as the LHC monitoring the patience, interest, knowledge, or some other property of the human being using the data server that the PC accesses.  Such reversibility raises the question of ``what is a system?'' rather forcefully: we humans understand what it means for another human to pick out the beam current in the LHC as a system of interest - at least we think we do - but have no understanding of what it might mean for the LHC to pick out a particular human being or that human being's apparent mental states as a system of interest.  Yet this is something that QBist agents do, and the LHC, like every other bounded system in the universe, is a QBist agent.  We humans also understand - again, we think we do - what it means for a human agent to have a belief, and to revise that belief given new data, but we do not understand what it might mean for the LHC to have a belief.  What is most odd, however, is the use that the two agents make of the universal ``users' manual'' of quantum mechanics.  The LHC, and indeed all physical systems on which measurements have been performed to date by us, appears to employ the manual perfectly; these systems appear to place their bets and take action in exact accord with the normative prescriptions of the Born rule, prescriptions which Fuchs insists any agent can ignore if they choose (p. 8, Fig. 2 caption).  Humans, on the other hand, are lousy at quantum mechanics.  Humans are notoriously bad at making probability assignments that cohere with available data and applying Bayes' theorem even in classical domains (e.g. \cite{kahneman-tversky:74, kahneman-tversky:81, pinker:97}).  For all of our pride in our comparatively large values of $d$, we seem to be dumber - at least, less coherent in our subjective probability assignments - when it comes to QBist quantum mechanics than electrons.

The binary view of measurement interactions that Fuchs illustrates in his Fig. 1, however, leaves out the agent with the largest value of $d$: the rest of the universe.  Fuchs explicitly considers the universe minus a particular agent to be a quantum system about which that agent can have quantum information, devoting all of Sect. VII to quantum cosmology.  The universe minus two agents must therefore be a quantum system as well, and hence be a third agent in its own right, conducting continuous measurements on the ``agent'' and ``system'' of any binary measurement interaction.  The rest of the universe, like electrons but unlike human beings, appears to be quite good at quantum mechanics, updating its beliefs about what it observes and acting on the basis of these beliefs, as far as we can tell, exactly in accord with the Born rule.  Just as it does in decoherence theory, the all-observing rest of the QBist universe acts as a witness to all binary measurement interactions.  Fuchs scolds an imagined colleague who asks what ``this world of inherent uncertainty'' is for God, saying ``trying to give \textit{him} a quantum state was what caused this trouble in the first place'' (p. 8; emphasis in original).  Giving a ``quantum state,'' i.e. quantum information and the users' manual to manipulate it, to the rest of the universe is, however, just as bad as giving it to God.  The consequences for QBism of the universe having access to quantum information are best understood by comparing the QBist situation to that in quantum Darwinism, where the ``environment as witness'' is explicitly embraced \cite{zurek:03rev, zurek:04, zurek:05, zurek:06, zurek:09rev}.

Let us return to the story of Wigner and his friend, and assume that Wigner's friend is effectively omniscient, at least about the present: he does not know what either the system or Wigner will do next, but he knows everything else, and performs every measurement continuously at the highest possible precision.  Under these conditions, the most efficient way for Wigner to discover information about the system of interest is to ask his friend.  This is what happens in quantum Darwinism: observers interact with their local fragments of the environment, which encode information about systems of interest.  For example, observers get information about the positions of distant objects by interacting locally with photons of ambient light that encode such positions.  Environmental encodings are massively redundant, and are assumed within the environment as witness framework to be separable; hence many observers can interact with their local environments simultaneously and arrive at a consensus about objective properties of systems of mutual interest as discussed above.  As is also discussed above, such consensus building requires both that the observers agree in advance as to what the system of mutual interest is and that they have system re-identification methods that commute with all other observables and work at least most of the time \cite{fields:10, fields:11}.  Advance agreement is required because the observers have to ask the environment the right questions: the environment simultaneously encodes the measureable properties of \textit{all possible} systems, and it does not, by assumption, know which of these systems the observers have agreed to be interested in unless they tell it.  Knowing what to ask the environment amounts, however, to knowing in advance how the environment encodes the states of every possible system of interest.  Standard presentations of the environment as witness formalism (\cite{zurek:04, zurek:05, zurek:06, zurek:07grand, zurek:09rev}; see also \cite{schloss:07, landsman:07, wallace:08}) miss this key point, as they treat observation as binary and define the environment relative to a single pre-specified system of interest.  If the ``environment'' is defined to be everything except, let us say, all simply-connected solid objects larger than 1 mm$^{3}$ within 1 km of the observer, the problem of how a quantum Darwinist observer knows what she is looking at leaps to the fore.

The situation is exactly the same in QBism.  The ``rest of the universe'' becomes Wigner's omniscient friend, who has assigned the most coherent probabilities possible to all possible outcomes, and keeps these probabilities updated continuously in exact accord with the Born rule.  Anything Wigner himself could measure could be found out, with better accuracy, simply by asking his friend.  The probabilities that the omniscient friend provides in response to questions are by assumption optimally current and optimally coherent, and so are ``objective'' probabilities in all but name, just the kind of probabilities that QBism rejects out of hand (p. 4-7).  Hence fully subjective probabilities can be saved, within QBism, only if it is assumed that observers \textit{cannot ask the environment} about the states of systems of interest.  This is the familiar QBist assumption of purely local interactions, but with the emphasis on the requirement that these interactions be \textit{directly} with the systems of interest, not mediated by the environment acting as a channel, and with the further ``taking for granted'' that QBist observers know what they are looking at.  The QBist environment is a witness to all interactions, but is a witness with whom QBist observers are forbidden, in principle, to communicate.

The fundamental paradox of the QBist observer now becomes clear: if QBist observers are physical systems, they cannot avoid interactions with the environment, and hence cannot avoid the environment's selection of experiences in reaction.  An observer can choose to ``ignore'' the environment, as in early decoherence theory, but cannot shield him/her/its-self from environmental interactions and their effects.  An observer could assume, \textit{ad hoc} and with no possibility of supporting evidence, that the experiences induced by the environment as a whole are identical to those induced by the system of interest, \textit{for every possible system of interest}.  This is an extraordinarily strong assumption, since the environment has by definition larger $d$ than either the observer or the system being observed, and in QBism $d$ measures \textit{autonomy}.  The assumption that the environment is in every case experientially transparent is thus an assumption that it is transparent in spite of its inherent unpredictability.  Absent such a seemingly unjustifiable assumption, the environment can be made to ``disappear'' in Fuchs' Fig. 1 by only one means: pulling the entire environment into the system of interest.  This solution is radical but effective: every ``real existence'' - in non-QBist terms, every collection of physical degrees of freedom - in the universe counts as a QBist observer, and every QBist observer interacts with and hence forms beliefs about exactly one system, the entire rest of his/her/its universe.  The problems of diachronic system identification and communication between observers vanish, as does the problem of accomodating fixed $d$.  This solution is \textit{not} solipsism; the rest of the universe is ``real'' and is clearly distinct from the observer.  It is, however, intensely personal.  QBist information is indeed information ``for \textit{me};'' besides the observer and his/her/its personal world, nothing else exists.

The true status of the QBist observer reveals, as anticipated, the true QBist ontology.  Personal probabilities plus local interactions yield a purely-personal world that is observer-dependent in the precise sense of comprising all of ``real existence'' - all physical degrees of freedom - not contained within the observer.  This world includes a single boundary, the observer-environment boundary, at which all interactions take place.  From the perspective of the observer, the environment is an autonomous agent and vice-versa.  This ontology thus appears to have all of the characteristics Fuchs desires.  They come, however, at a price.  The ``real world'' is maintained, but is rendered unrecognizable by the removal of all boundaries except the one around the observer.  This boundary-less world is as holist as Everett's world; indeed, it \textit{is} Everett's world, viewed from the perspective of a single observer.  In a universe in which all systems are agents, the price of personal probabilities is that the Einsteinian intuition of separability becomes untenable.

Once again, what holds for QBism holds for quantum Darwinism, but for different reasons.  Quantum Darwinist observers communicate \textit{only} with their local environments, i.e. only with their omniscient friends.  Observations of local encodings alone, however, are never sufficient to solve the reverse problem of determining what properties of what distant systems have been encoded locally, as demonstrated for the case of observers of classical finite automata by \cite{moore:56} over 50 years ago and as taken for granted in algorithmic studies of perception at least since the work of \cite{marr:82}.  Hence the boundaries of distant systems can never be determined by a quantum Darwinist observer; such an observer can assume separability, but this assumption has no operational meaning unless the observer invokes Zurek's ``axiom(o)'' to declare that each local encoding that is encountered is an encoding of some property of a particular re-identifiable bounded thing \cite{fields:10, fields:11}.  Like the QBist assumption of an experientially transparent environment, such axiomatic assumptions are \textit{ad hoc} and unverifiable, even in principle, and they rapidly approach utter incoherence when alternative local encodings specify alternative system boundaries and hence competing interpretations of what ``things'' are present in the world.  Without such object-specific axiomatic assumptions of ``thing-ness,'' the quantum Darwinist observer, like the QBist observer, interacts simply with ``the world'': all physical degrees of freedom of the universe other than those contained in him/her/its-self.

\section{Conclusion}

QBism appears at first glance to be, and is certainly presented by Fuchs as being, an entirely new approach to quantum foundations in which non-determinism is embraced as a virtue and the traditional foundational problems - in particular the problem of non-locality - simply disappear.  What has been shown here is that this is an illusion.  The foundational problems of quantum mechanics disappear in QBism only because Fuchs steadfastly refuses to follow through on the ``how does it work'' question, halting the explanation with the claim that systems select the POVM components that observers experience.  As soon as this question is considered, the traditional problems emerge, in some cases with even more virulence than they display in more standard approaches.  QBism provides no physical distinction between observers and the systems they observe, treating all quantum systems as autonomous agents that respond to observations by updating beliefs and employ quantum mechanics as a ``users' manual'' to guide behavior.  However, it treats observation itself as a physical process in which an ``observer'' acts on a ``system'' with a POVM and the ``system'' selects a POVM component as the ``observer's experience'' in return.  This requirement renders the assumption that systems be well-defined - i.e. have constant $d$ - impossible to implement operationally.  It similarly forces the consistent QBist to regard the environment as an effectively omniscient observer, threatening the fundamental assumption of subjective probabilities and forcing the conclusion that QBist observers cannot segment their environments into objectively separate systems.    

What is of even greater interest than the emergence of familiar problems in a new context, however, is that the virulent versions of these problems revealed by QBism can also be found lurking just behind their more familiar manifestations even in more traditional interpretative approaches, and in particular in quantum Darwinism.  As noted earlier, quantum Darwinism is about as far from QBism in terms of advertised assumptions as one can get.  If the present analysis is correct, QBism and quantum Darwinism share an apparently intractable problem with diachronic system identification, and are both driven to an ontological stance in which, absent indefensibly strong \textit{ad hoc} assumptions, the boundaries of external systems vanish, leaving the observer interacting with an undifferentiated ``personal universe.''  Shared deep problems suggest a shared implicit assumption, and the shared implicit assumption in this case appears to be that diachronic system identification is possible as a matter of objective, physical fact.  If this assumption is wrong, separability has no objective meaning.  Physics has come to accept and even to productively employ non-locality in the form of carefully-regulated entanglement \cite{nielsen-chaung:00, deutsch:85, galindo:01, gisin:05}; perhaps it can be formulated in a way that also accepts and derives insight from the notion of a universe in which separability plays no objective role.

\section*{Acknowledgements}

Thanks to Eric Dietrich and Ruth Kastner for helping to clarify my thinking about systems and their boundaries.  An earlier version of this paper was presented at the 2011 Boulder Conference on the History and Philosophy of Science; thanks to the audience at this conference for questions and comments.  Two anonymous referees also provided valuable comments on the manuscript.


\begin{thebibliography}{10}

\bibitem{fuchs:10}
C.~Fuchs.
\newblock Qbism: The frontier of quantum Bayesianism.
\newblock Preprint arXiv:1003.5209v1 [quant-ph], 2010.

\bibitem{fuchs:02}
C.~Fuchs.
\newblock Quantum mechanics as quantum information (and only a little more).
\newblock Preprint arXiv:quant-ph/0205039v1, 2002.

\bibitem{nielsen-chaung:00}
M.~A. Nielsen and I.~L. Chaung.
\newblock {\em Quantum Information and Quantum Computation}.
\newblock Cambridge University Press, Cambridge, 2000.

\bibitem{caves:02a}
C.~M. Caves, C.~A. Fuchs, and R.~Schack.
\newblock Quantum probabilities as Bayesian probabilities.
\newblock {\em Phys. Rev. A}, 65:022305, 2002.
\newblock arXiv:quant-ph/0106133v2.

\bibitem{caves:02b}
C.~M. Caves, C.~A. Fuchs, and R.~Schack.
\newblock Unknown quantum states: The quantum de Finetti representation.
\newblock {\em J. Math. Phys.}, 43:4537, 2002.
\newblock arXiv:quant-ph/0104088v1.

\bibitem{caves:07}
C.~M. Caves, C.~A. Fuchs, and R.~Schack.
\newblock Subjective probability and quantum certainty.
\newblock {\em Stud. Hist. Phil. Mod. Phys.}, 38:255--274, 2007.
\newblock arXiv:quant-ph/0608190v2.

\bibitem{palge:08}
V.~Palge and T.~Konrad.
\newblock A remark on fuch's Bayesian interpretation of quantum mechanics.
\newblock {\em Stud. Hist. Phil. Mod. Phys.}, 39:273--287, 2008.

\bibitem{timpson:08}
C.~G. Timpson.
\newblock Quantum Bayesianism: A study.
\newblock {\em Stud. Hist. Phil. Mod. Phys.}, 39:579--609, 2008.

\bibitem{friederich:11}
S.~Friederich.
\newblock How to spell out an epistemic conception of quantum states.
\newblock {\em Stud. Hist. Phil. Mod. Phys.}, 42:149--157, 2011.
\newblock arXiv:1101.1975v1 [quant-ph].

\bibitem{fuchs:11a}
C.~Fuchs and R.~Schack.
\newblock A quantum Bayesian route to quantum state space.
\newblock {\em Found. Phys.}, 41:345--356, 2011.
\newblock arXiv:0912.4252v1 [quant-ph].

\bibitem{cartwright:99}
N.~Cartwright.
\newblock {\em The Dappled World: A Study of the Boundaries of Science}.
\newblock Cambridge University Press, Cambridge, 1999.

\bibitem{bub:04}
J.~Bub.
\newblock Why the quantum.
\newblock {\em Stud. Hist. Phil. Mod. Phys.}, 35:241--266, 2004.

\bibitem{zurek:98rough}
W.~H. Zurek.
\newblock Decoherence, einselection and the existential interpretation (the
  rough guide).
\newblock {\em Phil. Trans. Royal Soc. A}, 356:1793--1821, 1998.

\bibitem{zurek:03rev}
W.~H. Zurek.
\newblock Deoherence, einselection, and the quantum origins of the classical.
\newblock {\em Rev. Mod. Phys.}, 75:715--775, 2003.
\newblock arXiv:quant-ph/0105127v3.

\bibitem{zurek:04}
H.~Ollivier, D.~Poulin, and W.~H. Zurek.
\newblock Objective properties from subjective quantum states: Environment as a
  witness.
\newblock {\em Phys. Rev. Lett.}, 93:220401, 2004.
\newblock arXiv:quant-ph/0307229v2.

\bibitem{zurek:05}
H.~Ollivier, D.~Poulin, and W.~H. Zurek.
\newblock Environment as a witness: Selective proliferation of information and
  emergence of objectivity in a quantum universe.
\newblock {\em Phys. Rev. A}, 72:042113, 2005.
\newblock arXiv:quant-ph/0408125v3.

\bibitem{zurek:06}
R.~Blume-Kohout and W.~H. Zurek.
\newblock Quantum Darwinism: Entanglement, branches, and the emergent
  classicality of redundantly stored quantum information.
\newblock {\em Phys. Rev. A}, 73:062310, 2006.
\newblock arXiv:quant-ph/0505031v2.

\bibitem{zurek:07grand}
W.~H. Zurek.
\newblock Relative states and the environment: Einselection, envariance,
  quantum Darwinism, and the existential interpretation.
\newblock Preprint arXiv:0707.2832v1 [quant-ph], 2007.

\bibitem{zurek:09rev}
W.~H. Zurek.
\newblock Quantum Darwinism.
\newblock {\em Nat. Phys.}, 5:181--188, 2009.
\newblock arXiv:0903.5082v1 [quant-ph].

\bibitem{joos-zeh:03}
E.~Joos, D.~Zeh, C.~Kiefer, D.~Giulini, J.~Kupsch, and I.-O. Stamatescu.
\newblock {\em Decoherence and the Appearance of a Classical World in Quantum
  Theory (2nd Ed.)}.
\newblock Springer, Berlin, 2003.
\newblock second edition.

\bibitem{schloss:04}
M.~Schlosshauer.
\newblock Decoherence, the measurement problem, and interpretations of quantum
  theory.
\newblock {\em Rev. Mod. Phys.}, 76:1267--1305, 2004.
\newblock arXiv:quant-ph/0312059v4.

\bibitem{schloss:07}
M.~Schlosshauer.
\newblock {\em Decoherence and the Quantum to Classical Transition}.
\newblock Springer, Berlin, 2007.

\bibitem{bohr:28}
N.~Bohr.
\newblock The quantum postulate and the recent developments of atomic theory.
\newblock {\em Rev. Mod. Phys.}, 121:580--590, 1928.

\bibitem{fuchs:11}
C.~Fuchs and R.~Schack.
\newblock Bayesian conditioning, the reflection principle and quantum
  decoherence.
\newblock Preprint arXiv:1103.5950v1 [quant-ph], 2011.

\bibitem{landauer:99}
R.~Landauer.
\newblock Information is a physical entity.
\newblock {\em Physica A}, 263:63--67, 1999.

\bibitem{griffiths:07}
R.~B. Griffiths.
\newblock Types of quantum information.
\newblock {\em Phys. Rev.}, 76:062320, 2007.
\newblock arXiv:0707.3752v2 [quant-ph].

\bibitem{mermin:98}
N.~D. Mermin.
\newblock The Ithaca interpretation of quantum mechanics.
\newblock {\em Pramana}, 51:549--565, 1998.

\bibitem{griffiths:11}
R.~B. Griffiths.
\newblock A consistent quantum ontology.
\newblock Preprint arXiv:1105.3932v1 [quant-ph], 2011.

\bibitem{everett:57}
H.~Everett.
\newblock `Relative state' formulation of quantum mechanics.
\newblock {\em Rev. Mod. Phys.}, 29:454--462, 1957.

\bibitem{tegmark:10}
M.~Tegmark.
\newblock Many worlds in context.
\newblock In S.~Saunders, J.~Barrett, A.~Kent, and D.~Wallace, editors, {\em
  Many Worlds? Everett, Quantum Theory and Reality}, pages 553--581. Oxford
  University Press, Oxford, 2010.
\newblock arXiv:0905.2182v2 [quant-ph].

\bibitem{schloss:06}
M.~Schlosshauer.
\newblock Experimental motivation and empirical consistency of minimal
  no-collapse quantum mechanics.
\newblock {\em Annals Phys.}, 321:112--149, 2006.
\newblock arXiv:quant-ph/0506199v3.

\bibitem{landsman:07}
N.~P. Landsman.
\newblock Between classical and quantum.
\newblock In J.~Butterfield and J.~Earman, editors, {\em Handbook of the
  Philosophy of Science: Philosophy of Physics}, pages 417--553. Elsevier,
  Amsterdam, 2007.
\newblock arXiv:quant-ph/0506082v2.

\bibitem{wallace:08}
D.~Wallace.
\newblock Philosophy of quantum mechanics.
\newblock In D.~Rickles, editor, {\em The Ashgate Companion to Contemporary
  Philosophy of Physics}, pages 16--98. Ashgate, Aldershot, 2008.
\newblock arXiv:0712.0149v1 [quant-ph].

\bibitem{conway-kochen:06}
J.~Conway and S.~Kochen.
\newblock The free will theorem.
\newblock {\em Found. Phys.}, 36:1441--1473, 2006.
\newblock arXiv:quant-ph/0604079v1.

\bibitem{conway-kochen:09}
J.~Conway and S.~Kochen.
\newblock The strong free will theorem.
\newblock {\em Notices AMS}, 56:226--232, 2009.
\newblock arXiv:0807.3286v1 [quant-ph].

\bibitem{bohr:13}
N.~Bohr.
\newblock On the constitution of atoms and molecules.
\newblock {\em Phil. Mag. (Ser. 6)}, 26:1--25, 1913.

\bibitem{fields:12}
C.~Fields.
\newblock If physics is an information science, what is an observer?
\newblock {\em Information}, 3:92--123, 2012.
\newblock arxiv:1108.4865 [quant-ph].

\bibitem{wallace:05}
D.~Wallace.
\newblock Everett and structure.
\newblock {\em Stud. Hist. Phil. Mod. Phys.}, 34:87--105, 2005.
\newblock arXiv:quant-ph/0107144v2.

\bibitem{sacks:85}
O.~Sacks.
\newblock {\em The Man Who Mistook His Wife for a Hat}.
\newblock Summit, New York, 1985.

\bibitem{fields:10}
C.~Fields.
\newblock Quantum Darwinism requires an extra-theoretical assumption of
  encoding redundancy.
\newblock {\em Int. J. Theor. Phys.}, 49:2523--2527, 2010.
\newblock arXiv:1003.5136v2 [quant-ph].

\bibitem{fields:11}
C.~Fields.
\newblock Classical system boundaries cannot be determined within quantum
  Darwinism.
\newblock {\em Phys. Essays}, 24:518--522, 2011.
\newblock arXiv:1008.0283v4 [quant-ph].

\bibitem{zeh:70}
D.~Zeh.
\newblock On the interpretation of measurement in quantum theory.
\newblock {\em Found. Phys.}, 1:69--76, 1970.

\bibitem{zeh:73}
D.~Zeh.
\newblock Toward a quantum theory of observation.
\newblock {\em Found. Phys.}, 3:109--116, 1973.

\bibitem{zurek:81}
W.~H. Zurek.
\newblock Pointer basis of the quantum apparatus: Into what mixture does the
  wave packet collapse?
\newblock {\em Phys. Rev. D}, 24:1516--1525, 1981.

\bibitem{zurek:82}
W.~H. Zurek.
\newblock Environment-induced superselection rules.
\newblock {\em Phys. Rev. D}, 26:1862--1880, 1982.

\bibitem{joos-zeh:85}
E.~Joos and D.~Zeh.
\newblock The emergence of classical properties through interaction with the
  environment.
\newblock {\em Z. Phys. B: Condensed Matter}, 59:223--243, 1985.

\bibitem{zeh:06}
D.~Zeh.
\newblock Roots and fruits of decoherence.
\newblock In B.~Duplantier, J.-M. Raimond, and V.~Rivasseau, editors, {\em
  Quantum Decoherence}, pages 151--175. Birkh{\"a}user, Basel, 2006.
\newblock arXiv:quant-ph/0512078v2.

\bibitem{vonNeumann:32}
J.~von Neumann.
\newblock {\em Mathematische Grundlagen der Quantenmechanik}.
\newblock Springer, Berlin, 1932.

\bibitem{kahneman-tversky:74}
A.~Tversky and D.~Kahneman.
\newblock Judgment under uncertainty: Heuristics and biases.
\newblock {\em Science}, 4157:1124--1131, 1974.

\bibitem{kahneman-tversky:81}
A.~Tversky and D.~Kahneman.
\newblock The framing of decisions and the psychology of choice.
\newblock {\em Science}, 4481:453--458, 1981.

\bibitem{pinker:97}
S.~Pinker.
\newblock {\em How the Mind Works}.
\newblock Norton, New York, 1997.

\bibitem{moore:56}
E.~F. Moore.
\newblock Gedankenexperiments on sequential machines.
\newblock In C.~W. Shannon and J.~McCarthy, editors, {\em Autonoma Studies},
  pages 129--155. Princeton University Press, Princeton, NJ, 1956.

\bibitem{marr:82}
D.~Marr.
\newblock {\em Vision}.
\newblock Freeman, New York, 1982.

\bibitem{deutsch:85}
D.~Deutsch.
\newblock Quantum theory, the Church-Turing principle and the universal quantum
  computer.
\newblock {\em Proc. Royal Soc. london A}, 400:97--117, 1985.

\bibitem{galindo:01}
A.~Galindo and M.~A. Martin-Delgado.
\newblock Information and computation: Classical and quantum aspects.
\newblock {\em Rev. Mod. Phys.}, 74:347--423, 2001.

\bibitem{gisin:05}
N.~Gisin.
\newblock Can relativity be considered complete? from Newtonian nonlocality to
  quantum nonlocality and beyond.
\newblock Preprint arXiv:quant-ph/0512168v1, 2005.

\end{thebibliography}
\end{document}